\documentclass[12pt]{article}
\usepackage[colorlinks=true,linkcolor=black,anchorcolor=black,citecolor=black,filecolor=black,menucolor=black,urlcolor=black]{hyperref}
\usepackage{amssymb,amsmath,amsfonts}
\usepackage{color}
\usepackage{latexsym}
\usepackage{graphicx}
\usepackage{mathrsfs}
\usepackage{dsfont}
\usepackage{textcomp}
\usepackage{graphicx}
\usepackage{amssymb}
\usepackage{hyperref}
\usepackage{rotating}

\newcommand{\be}{\begin{equation}}
\newcommand{\ee}{\end{equation}}
\newcommand{\bea}{\begin{eqnarray}}
\newcommand{\eea}{\end{eqnarray}}
\newcommand{\ba}{\begin{array}}
\newcommand{\ea}{\end{array}}

\newcommand{\hidden}[1]{}

\begin{document}

\begin{titlepage}
 \thispagestyle{empty}

 \begin{flushright}
 \end{flushright}

 \vspace{50pt}

\begin{center}
     { \huge{\bf      {Multi-Centered Black Hole Flows}}}

     \vspace{25pt}

     {\Large {Armen Yeranyan$^{a,b}$}}

     \vspace{20pt}

    {\it ${}^a$ INFN - Laboratori Nazionali di Frascati,\\
     Via Enrico Fermi 40, I-00044 Frascati, Italy\\
     \texttt{ayeran@lnf.infn.it}}

     \vspace{10pt}

{\it ${}^b$ Department of Physics, Yerevan State University\\
Alex Manoogian St. 1, Yerevan, 0025, Armenia}\\

     \vspace{10pt}

     \vspace{30pt}

\begin{abstract}

We describe the systematical construction of the first order formalism for
multi-centered black holes with flat three dimensional base-space, within
the so-called $T^{3}$ model of $N=2$, $D=4$ ungauged Maxwell-Einstein
supergravity.

The three possible flow classes (BPS, composite non-BPS and almost BPS) are
analyzed in detail, and various solutions, such as single-centered (static
or under-rotating) and all known multi-centered black holes, are recovered
in this unified framework. We also consider the possibility of obtaining new solutions.

The almost BPS class is proved to split into two general sub-classes,
corresponding to a positive or negative value of the duality-invariant
polynomial for the total charge; the well known almost BPS system is shown
to be a particular solution of the second sub-class.

\end{abstract}

 \vspace{10pt}
 \end{center}

\end{titlepage}



\section{Introduction}

Multi-centered non supersymmetric extremal black holes have been the subject
of a number of recent studies \cite{Bossard:2011kz}-\cite{Perz}, one possible
reason being that supersymmetry actually plays a less crucial r\^{o}le than
extremality in the investigation of these solutions.

The supersymmetric class for multi-centered black holes was investigated by
Denef \cite{Denef}, in order to match the true BPS spectrum and the
spectrum of spherically symmetric black holes in supergravity. The resulting
multi-centered BPS first order equations, whose solutions and properties
were then studied in \cite{Bates}, acquire the well-known form after
integrating out the auxiliary phase field in the single-centered limit.

More than ten years passed before a systematical construction of solutions
for the under-rotating non-BPS branch was worked out by exploiting
group-theoretical techniques \cite{Bossard:2011kz, Bossard:2012ge}.\medskip

The present paper is devoted to the determination of the most general flow
equations underlying the first order formalism \cite{Ceresole:2007wx}- \cite%
{Ceresole:2009iy} for non-BPS multi-centered and/or rotating black
holes with flat three dimensional base-space. For simplicity's sake, we will
consider the so-called $T^{3}$ model of $N=2$, $D=4$ ungauged
Maxwell-Einstein supergravity. But we expect all the results to be
generalizable to the $STU$ model, a task left to further forthcoming
investigations.

The first attempts to write down flow equations for non-supersymmetric multi-centered configurations were made in \cite{Perz}. The method was based on the squaring action procedure, but unfortunately an explicit form of flow equations in terms of charges, fields and other parameters was missing. In the present investigation, we try to fill this gap, and write down first order equations for non-supersymmetric multi-centered configurations in an explicit form.

Besides its inner simplicity and elegance, the importance first order
formalism relies in the possibility to switch from second order differential
equations of motion to the first order ones, without doubling their number.
Indeed, even if some auxiliary fields are added in the multi-centered case,
they turn out to satisfy algebraic equations. This is due to the fact that
the scalar charges are not independent, and the formalism automatically
discards the blowing up solutions. Integrating first order differential
equations is certainly easier, and thus the possibilities to find the
corresponding attractor flows are considerably enhanced. In this paper, we
will retrieve a number of known solutions in a unifying setting, and we will
predict the existence of other new ones, at least at the level of the most
general system of equations governing the flow.\medskip

The paper is organized as follows.

The setup and notations are defined in\ Sec. \ref{Setup}.

Then, Sec. \ref{BPS-Class} presents the general method of construction of
first order equations, considering the well known BPS class.

In Sec. \ref{Composite-nBPS-Class}, this method is then applied to the
construction of the first order formalism for the composite non-BPS class, and well known examples, such as single-centered static and
under-rotating black holes, and multi-centered electric
and magnetic solutions, are recovered.

Sec. \ref{Almost-BPS-Class} exploits the same procedure for the broadest
class of almost BPS multi-centered solutions. The existence of
two general sub-classes of first order flow equations is proved, and the
well known almost BPS system \cite{Goldstein:2008fq} is retrieved as a
particular solution of the second class.

Finally, Sec. \ref{Conclusion} contains a summary and an outlook of results.

\section{\label{Setup}Setup and Notations}


The bosonic part of the $N=2$, $D=4$ Maxwell-Einstein supergravity
Lagrangian density action for the so-called $T^{3}$ model reads
\begin{equation*}
\displaystyle\frac{\mathcal{L}}{\sqrt{\rule{0pt}{0.8em}-g}}=-\frac{1}{2}%
\,R+g^{\mu \nu }\,G\,\partial _{\mu }t\partial _{\nu }{\bar{t}}+\frac{1}{4}%
\,\mu _{\Lambda \Sigma }F_{\mu \nu }^{\Lambda }F^{\Sigma \,\mu \nu }+\frac{1%
}{4}\,\nu _{\Lambda \Sigma }F_{\mu \nu }^{\Lambda }\ast F^{\Sigma \,\mu \nu
},
\end{equation*}%
where $F^{\Lambda }=d\,A^{\Lambda }$ is the electromagnetic field, $G=-\frac{%
3}{(t-\bar{t})^{2}}$ is the metric on special K\"{a}hler symmetric scalar
manifold $SL(2,\mathbb{R})/U(1)$ , and the scalar-vector couplings are given
by
\begin{equation*}
\begin{array}{l}
\displaystyle\mu _{\Lambda \Sigma }=\frac{i}{4}\,\left(
\begin{array}{ll}
(t-{\bar{t}})\,(t^{2}+4t{\bar{t}}+{\bar{t}}^{2}) & -3(t^{2}-{\bar{t}}^{2}) \\%
[3mm]
-3(t^{2}-{\bar{t}}^{2}) & 6(t-{\bar{t}})%
\end{array}%
\right) ,\quad \nu _{\Lambda \Sigma }=\frac{1}{4}\,\left(
\begin{array}{ll}
-(t+{\bar{t}})^{3} & 3(t+{\bar{t}})^{2} \\[3mm]
3(t+{\bar{t}})^{2} & -12(t+{\bar{t}})%
\end{array}%
\right) .%
\end{array}%
\end{equation*}

The Ansatz of under-rotating multi-centered black hole (with flat
$D=3$ spatial slices), independently on the branches, reads
\begin{equation}
ds^{2}=g_{\mu \nu }dx^{\mu }dx^{\nu }=e^{2U}(dt+\omega
_{i}dx^{i})^{2}-e^{-2U}\delta _{ij}dx^{i}dx^{j},
\end{equation}%
where $i$ denotes the Cartesian indices, and $\mu ,\nu $ are space-time
indices. The functions $U$, $\omega_{i}$ and the complex modulus field $t=x-i\,y$,
where $x$ is axion and $y$ is dilaton, are solutions of the following
equations \cite{Breitenlohner:1987dg}:

\begin{equation}
\begin{array}{ll}
\begin{array}{l}
\displaystyle\delta ^{ij}\,\partial _{i}f_{j}=0,\qquad \delta
^{ij}\,\partial _{i}\left[ e^{-2U}M_{\alpha \beta }\left( \partial
_{j}b^{\beta }-f_{j}b^{\beta }\right) \right] =0,%
\end{array}
&  \\[0.8em]
\begin{array}{l}
\displaystyle G\,\partial _{(i}t\partial _{j)}{\bar{t}}+\partial
_{i}U\partial _{j}U+\frac{1}{4}\,e^{4U}f_{i}f_{j}=-\frac{1}{2}%
\,e^{-2U}\partial _{i}b^{\alpha }M_{\alpha \beta }\partial _{j}b^{\beta },%
\end{array}
&  \\[0.8em]
\begin{array}{l}
\displaystyle\delta ^{ij}\,\partial _{i}\partial _{j}U+\frac{1}{2}%
\,e^{4U}\delta ^{ij}f_{i}f_{j}=-\frac{1}{2}\,e^{-2U}\delta ^{ij}\partial
_{i}b^{\alpha }M_{\alpha \beta }\partial _{j}b^{\beta },%
\end{array}
&  \\[0.8em]
\begin{array}{l}
\displaystyle\delta ^{ij}\,\partial _{i}\left( G\partial _{j}t\right) -\frac{%
\partial G}{\partial {\bar{t}}}\,\delta ^{ij}\partial _{i}t\partial _{j}{%
\bar{t}}=\frac{1}{2}\,e^{-2U}\delta ^{ij}\partial _{i}b^{\alpha }\frac{%
\partial M_{\alpha \beta }}{\partial t}\,\partial _{j}b^{\beta },%
\end{array}
&  \\[0.8em]
&
\end{array}
\label{eom}
\end{equation}%
where $\alpha ,\beta $ are symplectic indices.

In the present framework, $f_{i}$ corresponds to the rotation, and it is
defined as follows:
\begin{equation}
f_{i}\equiv -\delta _{ij}\varepsilon ^{jkl}\partial _{k}\omega _{l}.
\end{equation}%
On the other hand, $b^{\alpha }$ denotes the electromagnetic potential. As
usual, we define the dual field strength $G_{\Lambda }=\partial \mathcal{L}/\partial
F^{\Lambda }$ {and the whole field strengths' symplectic vector $F^{\alpha
}=\left( F^{\Lambda },G_{\Lambda }\right) $, so that the electric
component of the field strength $F^{\alpha }$ reads}
\begin{equation}
F_{0i}^{\alpha }=-\partial _{i}b^{\alpha }.
\end{equation}%
As for $M_{\alpha \beta }$, it is a positive definite symplectic symmetric
matrix, which defines the single-centered black hole potential \cite%
{Ferrara:1997tw} and reads
\begin{equation*}
\begin{array}{l}
M_{\alpha \beta }=\left(
\begin{array}{cc}
\displaystyle\mu _{\Lambda \Sigma }+\nu _{\Lambda \Lambda ^{\prime }}\mu
^{\Lambda ^{\prime }\Sigma ^{\prime }}\nu _{\Sigma ^{\prime }\Sigma } & %
\displaystyle\nu _{\Lambda \Lambda ^{\prime }}\mu ^{\Lambda ^{\prime }\Sigma
} \\
\displaystyle\mu ^{\Lambda \Lambda ^{\prime }}\nu _{\Lambda ^{\prime }\Sigma
} & \displaystyle\mu ^{\Lambda \Sigma }%
\end{array}%
\right) ,%
\end{array}%
\end{equation*}%
where $\mu _{\Lambda \Sigma }\mu ^{\Sigma \Lambda ^{\prime }}\equiv \delta
_{\Lambda }^{\Lambda ^{\prime }}$.

A crucial point is the integration of the electromagnetic field equations.
In the following treatment, we will use the following Ansatz:
\begin{equation}
e^{-2U}M_{\alpha \beta }\partial _{i}b^{\beta }=\Omega _{\alpha \beta
}\hat{Q}^{\beta}_{i} ,  \label{Ans-1}
\end{equation}%
where $\hat{Q}^{\beta}_{i}=\left( \partial _{i}H^{\beta }+f_{i}b^{\beta }\right)$ and $\Omega _{\alpha \beta }$ is the bilinear symplectic invariant
structure:
\begin{equation*}
\begin{array}{l}
\Omega _{\alpha \beta }=\left(
\begin{array}{cc}
\displaystyle0 & -\displaystyle\delta _{\Lambda }^{\Sigma } \\
\displaystyle\delta _{\Sigma }^{\Lambda } & \displaystyle0%
\end{array}%
\right) ,%
\end{array}%
\end{equation*}%
and $H^{\beta }$ are harmonic functions containing only appropriate electric
and magnetic black hole charges (namely, fluxes of the 2-form field
strengths).

It is worth remarking here that one may generally write $\hat{Q}^{\beta}_{i}=\left( H^{\beta }_i+f_{i}b^{\beta }\right)$,  where $H^{\beta }_i$ is the dual of the three dimensional field strength (not necessarily harmonic), and $\delta^{j\,i}\partial_{j}H^{\beta }_i=0$. However, we will see below that (\ref{Ans-1}) is effective for all
classes of multi-centered black holes and models under consideration.

Substituting $\partial _{i}b^{\beta }$ given by (\ref{Ans-1}) back into
equations (\ref{eom}), one obtains

\begin{equation}
\begin{array}{ll}
\begin{array}{l}
\displaystyle\delta ^{ij}\,\partial _{i}f_{j}=0,\qquad \,e^{-2U}M_{\alpha
\beta }\partial _{i}b^{\beta }=\Omega _{\alpha \beta }\hat{Q}^{\beta}_{i} ,%
\end{array}
&  \\[0.8em]
\begin{array}{l}
\displaystyle G\,\partial _{(i}t\partial _{j)}{\bar{t}}+\partial
_{i}U\partial _{j}U+\frac{1}{4}\,e^{4U}f_{i}f_{j}=e^{2U}\hat{V}_{i\,j},%
\end{array}
&  \\[0.8em]
\begin{array}{l}
\displaystyle\delta ^{ij}\,\partial _{i}\partial _{j}U+\frac{1}{2}%
\,e^{4U}\delta ^{ij}f_{i}f_{j}=e^{2U}\delta ^{ij}\,\hat{V}_{i\,j},%
\end{array}
&  \\[0.8em]
\begin{array}{l}
\displaystyle\delta ^{ij}\,\partial _{i}\left( G\partial _{j}t\right) -\frac{%
\partial G}{\partial {\bar{t}}}\,\delta ^{ij}\partial _{i}t\partial _{j}{%
\bar{t}}=e^{2U}\delta ^{ij}\frac{\partial \hat{V}_{i\,j}}{\partial t}.%
\end{array}
&  \\[0.8em]
&
\end{array}
\label{eom1}
\end{equation}%
where%
\begin{equation}
\hat{V}_{i\,j}\equiv \hat{Z}_{(i}\,\hat{\bar{Z}}_{j)}+\hat{Z}_{t\,(i}\,\hat{%
\bar{Z}}_{\bar{t}\,j)},
\end{equation}%
is a generalized black hole potential, and ($v^{\alpha }$ denotes
the covariantly holomorphic section of special K\"{a}hler geometry
throughout)
\begin{equation}
\hat{Z}_{i}=\hat{Q}^{\alpha}_{i}\,\Omega
_{\alpha \beta }\,v^{\beta },\quad \hat{Z}_{t\,i}=E^{t}\,D_{t}\hat{Z}%
_{i}=\hat{Q}^{\alpha}_{i} \,\Omega
_{\alpha \beta }\,v_{t}^{\beta },  \label{prolong}
\end{equation}%
are Cartesian-indexed generalization of the $N=2$ cental charge function and
of its first K\"{a}hler-covariant derivative
\begin{equation}
Z=Q^{\alpha}\,\Omega
_{\alpha \beta }\,v^{\beta },\quad Z_{t}=E^{t}\,D_{t}Z=Q^{\alpha} \,\Omega
_{\alpha \beta }\,v_{t}^{\beta },  \label{usual}
\end{equation}%
in which the charge vector
$Q^{\alpha }=\{p^{\Lambda },q_{\Lambda }\}$ is replaced by $\hat{Q}^{\alpha}_{i}$. $E^{t}\equiv E_{\widehat{t}}^{t}$ is the (inverse) Vielbein of the
scalar manifold ($\left\vert E^{t}\right\vert ^{-2}=G$); and we henceforth
refrain from hatting the subscript \textquotedblleft $t$", denoting the
flattened scalar coordinate.

Is worth mentioning here that an object like the generalized black hole potential $\hat{V}_{i\,j}$ appeared in \cite{Van}, but our definition has a more transparent geometrical interpretation.
 As for single center static and spherically symmetric black holes, only the $V_{\tau\, \tau}$ (where $\tau=1/r$ is the inverse radial coordinate) component survives in this case, and it equals the usual black hole potential $V_{BH}$ \cite{Ferrara:1997tw}.

\smallskip\

We are now going to construct the first order flows for these equations in
various branches. After \cite{Bossard:2011kz}, it is known that for
under-rotating multi-centered black holes three possible flow classes
exist, namely: BPS, non-BPS composite and almost BPS.

For all classes of multi-centered flows based on the known solutions \cite{Bossard:2011kz},\cite{Bossard:2012ge} and \cite{Denef}, one can consider the following Ansatz for the first order equations:
\begin{eqnarray}
\partial _{i}U &=&\hat{W}_{i}\left(\hat{Z}_{i},\,\hat{Z}_{t\,i},\,f_i,\alpha_a\right)e^{U}, \\
\partial _{i}t &=&E^{t}\,\hat{\Pi}_{i}\left(\hat{Z}_{i},\,\hat{Z}_{t\,i},\,f_i,\alpha_a\right)e^{U},\\
\partial _{i}\alpha_a &=&\Gamma_{i\,a}\left(\hat{Z}_{i},\,\hat{Z}_{t\,i},\,f_i,\alpha_a\right)e^{U}.
\end{eqnarray}%
Here $\hat{W}_{i}$ denotes a Cartesian-indexed generalization of the
first order (fake) superpotential $W$ for single center black hole [11]. $\hat{\Pi}_{i}$ is the
corresponding generalization of the scalar charges ($2 \bar{E}^{\bar{t}}\,\partial_{\bar t} W$), and $\alpha$'s are some auxiliary fields. \textit{\c{C}a va sans
dire}, these functions will generally be different for the various classes.

For pedagogical
reasons, we will start by considering the BPS class.

\section{\label{BPS-Class}BPS Flow}

This is a well known case \cite{Denef}, and we will consider it in order to
establish the procedure which we will exploit for the non-supersymmetric
cases.

For single-centered BPS black holes in spherical coordinates, the first
order superpotential reads
\begin{equation}
W=\sqrt{Z\,\bar{Z}}.  \label{W}
\end{equation}%
In order to re-express it in Cartesian coordinates, a mere replacement of
charges by derivative of harmonic functions (namely, $Z\rightarrow Z_{i}$)
is not convenient, because Cartesian indices would occur under square root,
and covariance would be lost. Actually, manifest covariance can be preserved
by introducing the phase $\alpha _{0}$ and expressing $W$ (\ref{W}) as \cite%
{Denef}

\begin{equation}
W=\text{Re}\left[e^{-i\alpha_0}\,Z\right],\quad e^{-i\,\alpha_0}=\sqrt{\frac{\bar{Z}%
}{Z}}.
\end{equation}

Then, one can easily switch to Cartesian coordinates by introducing the
corresponding index:%
\begin{equation}
W_{i}=\text{Re}\left[ e^{-i\alpha _{0}}\,Z_{i}\right] .
\end{equation}
Of course, the definition of $\alpha _{0}$ through $Z$ and $\bar{Z}$ does
not hold true anymore here, and the corresponding equation should be
established. Rotation can be introduced into by prolonging central charge
as $Z_{i}\rightarrow \hat{Z}_{i}$ (recall (\ref{prolong})). By exploiting
such a procedure also for the scalar charges, one ends up with the
following first order system:
\begin{eqnarray}
\partial _{i}U &=&\text{Re}\left[ e^{-i\alpha _{0}}\,\hat{Z}_{i}\right] e^{U}, \\
\partial _{i}t &=&E^{t}\,e^{i\,\alpha _{0}}\,\hat{\bar{Z}}_{\bar{t}%
\,i}\,e^{U}.
\end{eqnarray}%
By plugging these equations back into (\ref{eom1}), one then gets equations
for $\alpha _{0}$ and $f_{i}$, namely:
\begin{eqnarray}
\partial _{i}\alpha _{0} &=&e^{U}\,\text{Im}\left[ e^{-i\,\alpha _{0}}\left( \hat{Z}%
_{i}+\sqrt{3}\,\hat{Z}_{t\,i}\right) \right] , \\
f_{i} &=&2\text{Im}\left[ e^{-i\alpha _{0}}\,\hat{Z}_{i}\right] e^{-U}.  \label{16}
\end{eqnarray}

We notice that Eq. (\ref{16}) contains $f_{i}$ in both sides; thus, by
recalling definition (\ref{prolong}), it can be rewritten as follows:
\begin{equation}
(1-2\text{Im}\left[ e^{-i\alpha _{0}}\,b^{\alpha }\,\Omega _{\alpha \beta
}\,v^{\beta }\right] e^{-U})f_{i}=2\text{Im}\left[ e^{-i\alpha _{0}}\,Z_{i}\right]
e^{-U}.
\end{equation}

It is here worth noting that, due to the presence of the rotation, all
equations contain the electromagnetic potential. One can easily show that the
following Ansatz solves the equation for the electromagnetic potential
in a model-independent way:
\begin{equation}
b^{\beta }=2\text{Re}\left[ e^{-i\alpha _{0}}\,v^{\beta }\right] e^{U}.  \label{18}
\end{equation}%
We point out that this Ansatz can be constructed by replacing all $%
\hat{Z}_{i}$'s with the symplectic sections $v^{\beta }${\ in }the
superpotential $W_{i}$, {and then by multiplying by $2e^{U}$}. In the
following sections, we will see this procedure to (partially) work also
within non-supersymmetric classes of flows.

Plugging (\ref{18}) back into all other equations, one finally gets the
Denef system \cite{Denef}:

\begin{eqnarray}  \label{BPSsyst}
\partial _{i}U &=&\text{Re}\left[ e^{-i\alpha _{0}}\,Z_{i}\right] e^{U},
\label{BPS} \\
\partial _{i}t &=&E^{t}\,e^{i\,\alpha _{0}}\,\bar{Z}_{\bar{t}\,i}\,e^{U}, \\
\partial _{i}\alpha _{0} &=&-\text{Im}\left[ \frac{\partial K}{\partial t}%
\,\partial _{i}t\right] +\frac{1}{2}f_{i}\,e^{2U}, \\
f_{i} &=&-2\text{Im}\left[ e^{-i\alpha _{0}}\,Z_{i}\right] e^{-U},  \label{BPS-2}
\end{eqnarray}
where $K=-\log\left(-i(t-\bar{t})^3\right)$ is the K\"{a}hler potential.

The rotation should be divergence-free (namely, $\delta ^{ij}\,\partial
_{i}f_{j}=0$), thus so from (\ref{BPS-2}) one obtains
\begin{equation}
f_{i}=\partial _{i}H^{\alpha }\,\Omega _{\alpha \,\beta }\,H^{\beta }.
\label{fbps}
\end{equation}%
\medskip

Let us now consider the static single-centered limit of the system (\ref{BPS}%
)-(\ref{BPS-2}). By setting $f_{i}=0$ from the start, the standard BPS
solution is obtained without any integration of the equation of
electromagnetic fields; however, the condition
\begin{equation}
Q^{\alpha }\,\Omega _{\alpha \,\beta }\,h^{\beta }=0,  \label{ccond}
\end{equation}%
where $h^{\beta }$ are the asymptotical values of the harmonic functions $%
H^{\beta }$, should be introduced \textit{ad hoc}.

On the other hand, after integrating electromagnetic fields out and finding (%
\ref{fbps}), by setting $f_{i}=0$ one obtains the condition (\ref{ccond})
automatically.

\section{\label{Composite-nBPS-Class}Composite non-BPS Flow}

\subsection{General Analysis}

In this section we are going to investigate the so-called composite
non-BPS branch \cite{Bossard:2011kz}. In order to determine the fake
super potential and the corresponding scalar charges, in this case one can
start from the single-centered non-BPS black hole fake superpotential
\cite{Ceresole:2009iy}, and introduce one or more phases. Due to
the rather cumbersome explicit expression of the non-BPS $W$, it is not easy
to guess how to consistently introduce phases, so we choose to proceed in
another way. Namely, we will use the procedure described for the $N=2$, $D=4$
$STU$ model in \cite{Bossard:2011kz}. As treated in that paper,
under-rotating solutions correspond to various nilpotent orbits, and thus
one can try to switch from the BPS orbit to the non-BPS ones by
performing the following steps (for the $T^{3}$ model):

\begin{enumerate}
\item Consider  the scalar charges and the quantity whose real part is the BPS
superpotential $W=\text{Re}(w)$, and remove the phase $\alpha _{0}$
therein:
\begin{equation}
w=Z,\quad \Pi =\bar{Z}_{\bar{t}}.
\end{equation}

\item Switch to the duality frame (on spatial infinity) in which the axion $%
x $ vanishes and the dilaton $y$ is set to $1$, thus establishing a one to
one correspondence between the the charges and the central charge and its K%
\"{a}hler-covariant derivative. In such a way, all quantities can be
rewritten solely in terms of the charges themselves.

\item For the composite non-BPS orbit, the signs of $q_{0}$ and
of the combination $p^{0}+q_{1}/3$ should be flipped, and such quantities
should be written in terms of the central charge and of its derivative, thus
yielding:
\begin{equation}
w=\frac{1}{2}\left( \bar{Z}-\sqrt{3}\,\bar{Z}_{\bar{t}}\right) ,\quad \Pi =-%
\frac{1}{2}\,\left( \sqrt{3}\,Z+Z_{t}\right) .
\end{equation}

\item Then, the phases should be introduced using the following rules under $%
\left[ U(1)\right] ^{2}$ transformations \cite{Bossard:2011kz}:
\begin{equation}
Z\rightarrow Ze^{\frac{i}{2}(\alpha _{0}+3\,\alpha _{1})},\quad
Z_{t}\rightarrow Z_{t}e^{-\frac{i}{2}(\alpha _{0}+\alpha _{1})},\quad
w\rightarrow w\,e^{-\frac{i}{2}(\alpha _{0}+3\,\alpha _{1})},\quad \Pi
\rightarrow \Pi \,e^{\frac{i}{2}(\alpha _{0}-\alpha _{1})},
\end{equation}%
thus achieving the expressions for the fake superpotential and
corresponding scalar charges:
\begin{eqnarray}
W &=&\frac{1}{2}\text{Re}\left[ e^{-3i\alpha _{1}}\,\bar{Z}-\sqrt{3}e^{-i\alpha
_{1}}\,\bar{Z}_{\bar{t}}\right] ,  \label{28} \\
\Pi &=&-\frac{1}{2}\left( \sqrt{3}\,e^{i\alpha _{1}}\,Z+e^{-i\alpha
_{1}}\,Z_{t}\right) .  \label{29}
\end{eqnarray}
\end{enumerate}

Clearly, the expressions (\ref{28}) and (\ref{29}) are written only at
spatial infinity. However, we choose to use them all along the flow, by
just replacing the charges $Q^{\alpha }$ with the quantity $\hat{Q}^{\alpha }_i$, exhibiting the same symplectic
covariance.This replacement can be understood by observing that $\hat{Q}^{\alpha }_i$ naturally occurs in the second order equations (\ref{eom1}), and at infinity it equals $Q^{\alpha }$ when spherical symmetry is restored. After integrating the phase $\alpha _{1}$ out, we will show
below that the known expression of non-BPS $W$ for the $T^{3}$ model \cite%
{Ceresole:2009iy} is retrieved.

By using the expressions of $\hat{W}_{i}$ and $\hat{\Pi}_{i}$, which as in
the BPS class contain only one phase (in this case denoted as $\alpha _{1}$%
), the following first order equations can be written down:
\begin{eqnarray}
\partial _{i}U &=&\frac{1}{2}\text{Re}\left[ e^{-3i\alpha _{1}}\,\hat{\bar{Z}}_{i}-%
\sqrt{3}e^{-i\alpha _{1}}\,\hat{\bar{Z}}_{\bar{t}\,i}\right] e^{U}, \\
\partial _{i}t &=&-\frac{1}{2}E^{t}\,\left( \sqrt{3}\,e^{i\alpha _{1}}\,\hat{%
Z}_{i}+e^{-i\alpha _{1}}\,\hat{Z}_{t\,i}\right) \,e^{U}.
\end{eqnarray}%
Thence, by plugging them back into the second order equations (\ref{eom1}),
one obtains
\begin{eqnarray}
\partial _{i}\alpha _{1} &=&-\frac{2}{\sqrt{3}}e^{U}\,\sin {\alpha _{1}}\,\text{Re}%
\left[ e^{i\alpha _{1}}\,\hat{\Pi}_{i}\right] ,  \label{pre-fcomp} \\
f_{i} &=&\text{Im}\left[ e^{-3i\alpha _{1}}\,\hat{\bar{Z}}_{i}-\sqrt{3}e^{-i\alpha
_{1}}\,\hat{\bar{Z}}_{\bar{t}\,i}\right] e^{-U}.  \label{fcomp}
\end{eqnarray}%
It is here worth stressing that, as in the BPS class, $f_{i}$ is the
imaginary part of the quantity whose real part is $W_{i}$.

Moreover, $f_{i}$ enters both sides of (\ref{fcomp}); thus, we would like to
integrate out the electromagnetic potential completely, as for the
BPS class. Following the procedure outlined in Sec. \ref{BPS-Class}, we
replace the central charge and its flat derivative respectively by $%
v^{\alpha }$ and $v_{t}^{\alpha }$, obtaining the following Ansatz:
\begin{equation}
b^{\beta }=\text{Re}\left[ e^{-3i\alpha _{1}}\,\bar{v}^{\beta }-\sqrt{3}e^{-i\alpha
_{1}}\,\bar{v}_{\bar{t}}^{\beta }\right] e^{U}.  \label{Ans-not}
\end{equation}%
Unfortunately, this Ansatz does not work here well. Indeed, by
inserting it into (\ref{fcomp}), it turns out that $f_{i}$ drops out, and
one ends up with the expression:
\begin{equation}
\text{Im}\left[ e^{-3i\alpha _{1}}\,\bar{Z}_{i}-\sqrt{3}e^{-i\alpha _{1}}\,\bar{Z}_{%
\bar{t}\,i}\right] =0.  \label{concomp}
\end{equation}

{However, we can add to $b^{\beta }$ (\ref{Ans-not}) a new term proportional
to a boundary function $b$.} As we will see, $b$ represents a real
rotation, whereas in the BPS class only intrinsic rotation $f_{i}$ is
present. Consequently, the new Ansatz reads
\begin{equation}
b^{\beta }=\text{Re}\left[ e^{-3i\alpha _{1}}\,\bar{v}^{\beta }-\sqrt{3}e^{-i\alpha
_{1}}\,\bar{v}_{t}^{\beta }\right] e^{U}-b\,e^{3U}\text{Im}\left[ e^{-3i\alpha
_{1}}\,\bar{v}^{\beta }-\sqrt{3}e^{-i\alpha _{1}}\,\bar{v}_{t}^{\beta }%
\right] .  \label{concomp-2}
\end{equation}%
The new term added, proportional to $b$, is nothing but $f_{i}$ (\ref{fcomp}%
)(namely, the imaginary part of the quantity whose real part is $W_{i}$)
with the central charge and its flat derivative respectively replaced by $%
v^{\beta }$ and $v_{t}^{\beta }$. Using this Ansatz, one finds that
(\ref{concomp}) holds true, whereas the equation for electromagnetic fields
yields the following expression for $f_{i}$:
\begin{equation}
f_{i}=\partial _{i}b+2e^{-U}\text{Im}\left[ e^{-3i\alpha _{1}}\bar{Z}_{i}\right] +%
\sqrt{3}\,b\,e^{U}\text{Re}\left[ \sqrt{3}e^{3i\alpha _{1}}\,Z_{i}-e^{i\alpha
_{1}}\,Z_{t\,i}\right] .  \label{compf}
\end{equation}%
Finally, by inserting (\ref{concomp}) and (\ref{concomp-2}) back into the
other equations, one obtains full-fledged first order formalism for
the composite non-BPS class of under-rotating multi-centered
flows:
\begin{eqnarray}
\partial _{i}U &=&\frac{1}{2}\text{Re}\left[ e^{-3i\alpha _{1}}\,\bar{Z}_{i}-\sqrt{3%
}e^{-i\alpha _{1}}\,\bar{Z}_{\bar{t}\,i}\right] e^{U}+\frac{1}{2}%
\,b\,f_{i}\,e^{4U}  \label{sys-1} \\
\partial _{i}t &=&-\frac{1}{2}e^{U-2i\alpha _{1}}E^{t}\left( \sqrt{3}%
\,e^{U}\,\left( i-be^{2U}\right) f_{i}+\sqrt{3}\,e^{3i\alpha
_{1}}\,Z_{i}+e^{i\alpha _{1}}\,Z_{t\,i}\right) .  \label{sys-2}
\end{eqnarray}

\subsection{\label{Ex-1}Some Examples}

In order to find explicit solutions, the system (\ref{sys-1})-(\ref{sys-2})
should be supplemented by the condition (\ref{concomp}),which is the
integrated version of equation (\ref{pre-fcomp}).

\begin{enumerate}
\item Let us consider the single-centered non-rotating non-supersymmetic
black hole in the $T^{3}$ model. By setting $f_{i}=0$ before the second
integration of the electromagnetic field equations and bearing in mind that, by
virtue of spherical symmetry, all three Cartesian equations in (\ref{concomp}%
) can be replaced by the only radial coordinate, one obtains the following
solution for $\alpha _{1}$:%
\begin{eqnarray}
e^{2i\,\alpha _{1}} &=&-\frac{3\lambda +i\,\sqrt{-9\,\lambda
^{2}+24\,\lambda \,Z_{t}\,\bar{Z}_{\bar{t}}+16Z_{t}\,\bar{Z}_{\bar{t}%
}\,(3\,Z\,\bar{Z}-Z_{t}\,\bar{Z}_{\bar{t}})}}{4\,\sqrt{3}\,Z\bar{Z}_{\bar{t}}%
}+\frac{Z_{t}}{\sqrt{3}\,Z},  \label{al1} \\
\lambda &=&\left( \left( Z\bar{Z}-\frac{Z_{t}\bar{Z}_{\bar{t}}}{3}\right)
^{3}-I_{4}\,\left( Z\bar{Z}+Z_{t}\bar{Z}_{\bar{t}}\right) +\frac{4}{3\sqrt{3}%
}i\,(\bar{Z}Z_{t}^{3}-Z\bar{Z}_{\bar{t}}^{3})\,\sqrt{-I_{4}}\right) ^{1/3}+
\notag \\
&+&\left( \left( Z\bar{Z}-\frac{Z_{t}\bar{Z}_{\bar{t}}}{3}\right)
^{3}-I_{4}\,\left( Z\bar{Z}+Z_{t}\bar{Z}_{\bar{t}}\right) -\frac{4}{3\sqrt{3}%
}i\,(\bar{Z}Z_{t}^{3}-Z\bar{Z}_{\bar{t}}^{3})\,\sqrt{-I_{4}}\right) ^{1/3},
\end{eqnarray}%
where $I_{4}$ is quartic invariant. Plugging (\ref{al1}) back into
multi-centered fake superpotential (\ref{28}), the known expression for
the non-BPS fake superpotential of the $T^{3}$ model \cite{Ceresole:2009iy}
is retrieved:
\begin{equation}
W^{2}=\frac{1}{4}\left( Z\bar{Z}+Z_{t}\bar{Z}_{\bar{t}}\right) +\frac{3}{8}%
\lambda .
\end{equation}%
Of course, one can also consider the single-centered limit after the second
integration of electromagnetic potentials. In this case, one has to impose the
condition $f_{i}=0$, where $f_{i}$ is defined by expression (\ref{compf}).
By setting $b$ to a constant, and integrating $\alpha _{1}$ out, one
achieves the following condition:
\begin{equation}
\left( x^{2}+y^{2}\right) \left( -1+\nu ^{2}\right) -\sigma _{-}^{2}+\nu
^{2}\sigma _{+}^{2}+2be^{2u}y\nu (\sigma _{-}+\sigma _{+})-2x\left( \sigma
_{-}+\nu ^{2}\sigma _{+}\right) =0,  \label{conddd}
\end{equation}%
where $x\equiv \text{Re}\left( t\right) $ is the axion field, $y\equiv -\text{Im}\left(
t\right) $ is the dilaton field, and $\sigma _{+}$, $\sigma _{-}$ and $\nu $
are defined in formula (3.19) of \cite{Ceresole:2009iy}. It is easy to check
that solutions (4.19) of \cite{Ceresole:2009iy} indeed satisfy this
condition. {As for the BPS case, one obtains the condition (\ref{conddd})
in order to have no rotation, whereas in the previous approach (namely,
before integrating electromagnetic potential out) such condition had to be
imposed by hand.} Thus, one can conclude that the composite non-BPS class exhibits a smooth single-centered non-rotating non-BPS (extremal)
limit.

\item The example concerns the generalization of Rasheed-Larsen black hole
\cite{Rasheed:1995zv}, namely a non-BPS single-centered rotating black hole
with all charges switched on. As we are considering a single-centered black
hole, all harmonic functions sourced by the charges depend only on the radial
coordinate $r$. Thus, condition (\ref{concomp}) can be solved as in the case
of a non-rotating single-center black hole. By computing $\alpha _{1}$ and
inserting it back into the expressions for the derivatives of warp factor
and for scalar fields, one obtains the very same equations for the
non-rotating single-centered black hole, and the very same solutions (see
e.g. (4.19) of \cite{Ceresole:2009iy}), the unique difference being
that, b now is not constant anymore, but instead it is a harmonic function
which represents rotation:
\begin{equation}
b=b_{0}+J\frac{\cos {\theta }}{r^{2}},
\end{equation}%
where $b_{0}$, $J$ are constants and $J$ is the angular momentum. By setting
all charges to zero but the $D0$ and $D6$ ones (which, in the notation of
\cite{Ceresole:2009iy} amounts to setting $\nu \rightarrow 0$, $\sigma
_{-}\rightarrow 0$ and $\nu \,\sigma _{+}\rightarrow (q_0)^{1/3}/(p^0)^{1/3}$),
and by imposing the additional condition $h_{1}=h_{0}$ (corresponding to the
vanishing of the axion), the Rasheed-Larsen solution is consistently
reproduced.

\item As third example, we consider the ( $T^{3}$%
-degeneration of the) composite multi-centered black holes investigated by
Bossard and Ruef in \cite{Bossard:2011kz}, \cite{Bossard:2012ge}. In the
multi-centered case, the condition (\ref{concomp}) can explicitly be solved
only in some particular charge configurations. Namely, when all D6 charges
or all D0 charges vanish on all centers (respectively called
magnetic and electric configurations). For the magnetic configuration,
condition (\ref{concomp}) yields $\alpha _{1}=0$. By solving all other
equations, one gets the following solutions for warp factor, axion and
dilaton, respectively:
\begin{eqnarray}
U &=&-\frac{1}{4}\log {\left[ 4\,A\,(-H_{p^{1}})^{3}-b^{2}\right] }, \\
x &=&B+\frac{b}{2(H_{p^{1}})^{2}}, \\
y &=&\frac{e^{-2U}}{2(H_{p^{1}})^{2}}.
\end{eqnarray}%
Here $H_{p^{1}}<0$ denotes the harmonic function which corresponds to the
magnetic charges $p^{1}$, while the harmonic function $B$ satisfies
\begin{equation}
\partial _{i}B=\frac{6\,B\,\partial _{i}H_{p^{1}}-\partial _{i}H_{q_{1}}}{%
H_{p^{1}}}.
\end{equation}%
Furthermore, the non-harmonic function $A$ is constrained to satisfy the
equation%
\begin{equation*}
\partial _{i}A=\partial _{i}H_{q_{0}}+3\,A\,\frac{\partial _{i}H_{p^{1}}}{%
H_{p^{1}}}+B\,\partial _{i}H_{q_{1}}-3\,B^{2}\,\partial _{i}H_{p^{1}}.
\end{equation*}%
It is here worth mentioning that \ one could have found this solution
more easily by considering equations for the combinations $e^{-U}/\sqrt{y}$
and $x-y\,b\,e^{2U}$. As for $f_{i}$, it reads
\begin{equation}
f_{i}=\partial _{i}b+(H_{p^{1}})^{2}\,\partial _{i}B,
\end{equation}%
and $b$ is such that implies $f_{i}$ to be divergence-free. For the
electric configuration, condition (\ref{concomp}) yields $\alpha
_{1}=\text{arccot}\,\left( x/y\right) $ and the following solutions can be
obtained:
\begin{eqnarray}
U &=&-\frac{1}{4}\log {\left[ \frac{4}{27}\,A\,(H_{q_{1}})^{3}-b^{2}\right] }%
, \\
x &=&\frac{b-2/9(H_{q_{1}})^{2}B}{2\left(
-bB+(((H_{q_{1}})^{2}B^{2})/9+(H_{q_{1}}A)/3\right) }, \\
y &=&\frac{e^{-2U}}{2\left(
-bB+(((H_{q_{1}})^{2}B^{2})/9+(H_{q_{1}}A)/3\right) },
\end{eqnarray}%
where functions $B$ (harmonic) and $A$ (non-harmonic) satisfy:
\begin{eqnarray}
\partial _{i}B &=&\frac{2\,\partial _{i}H_{q_{1}}\,B-3\,\partial _{i}H_{p^{1}}}{%
2H_{q_{1}}}, \\
\partial _{i}A &=&\partial _{i}H_{p^{0}}+B\,(-3\,\partial
_{i}H_{p^{1}}+B\partial _{i}H_{q_{1}}),
\end{eqnarray}%
and $f_{i}$ is given by
\begin{equation}
f_{i}=\partial _{i}b-\frac{2}{3}\,(H_{q_{1}})^{2}\,\partial _{i}B.
\end{equation}

Since we are mostly interested in the construction of flow equations, we leave the analysis  of the physical properties of these solutions for further future investigation.
\end{enumerate}

\section{\label{Almost-BPS-Class}Almost BPS Flow}

\subsection{General Analysis}

Finally, let us investigate the almost BPS class.

This class can be obtained from the BPS one by exploiting the very same
procedure considered for the composite non-BPS class, the unique
difference being the fact that in this case only the sign of the electric
graviphoton charge is flipped: $q_{0}\rightarrow -q_{0}$. This yields the
following first order system:
\begin{eqnarray}
\partial _{i}U &=&\frac{1}{4}\,\text{Re}\left[ \left( 3\,e^{-i\,\alpha
_{0}}-e^{3\,i\,\alpha _{1}}\right) \,\hat{Z}_{i}-\sqrt{3}\,e^{-i\,(\alpha
_{0}+2\,\alpha _{1})}\left( 1+e^{i\,(\alpha _{0}+3\,\alpha _{1})}\right) \,%
\hat{Z}_{t\,i}\right] e^{U},  \label{almost} \\
\partial _{i}t &=&-\frac{1}{4}\,E^{t}\,\left( \sqrt{3}e^{i\,\alpha _{1}}\hat{%
Z}_{i}+\sqrt{3}e^{i\,(\alpha _{0}-2\alpha _{1})}\hat{\bar{Z}}%
_{i}+3e^{-i\,\alpha _{1}}\hat{Z}_{t\,i}-e^{i\,\alpha _{0}}\hat{\bar{Z}}_{%
\bar{t}\,i}\right) \,e^{U}.  \label{almost-2}
\end{eqnarray}%
By inserting (\ref{almost}) and (\ref{almost-2}) into the second order
equations (\ref{eom1}), a system of equations for the phases $\alpha _{0}$
and $\alpha _{1}$ and for $f_{i}$ is obtained:
\begin{eqnarray}
f_{i} &=&\frac{1}{2}\,\text{Im}\left[ \left( 3\,e^{-i\,\alpha _{0}}+e^{3\,i\,\alpha
_{1}}\right) \,\hat{Z}_{i}-\sqrt{3}\,e^{-i\,(\alpha _{0}+2\,\alpha
_{1})}\left( 1-e^{i\,(\alpha _{0}+3\,\alpha _{1})}\right) \,\hat{Z}_{t\,i}%
\right] e^{-U},  \label{almosfal} \\
\partial _{i}\alpha _{0} &=&\frac{\sqrt{3}}{2}\,e^{U}\,\text{Im}\left[
e^{-2i\,\alpha _{1}}\left( -1+2e^{2i\,\alpha _{1}}+e^{i(\alpha
_{0}+3\,\alpha _{1})}\right) \hat{\bar{\Pi}}_{i}\right] +  \notag \\
&+&\frac{1}{2}\,e^{U}\,\sin {\ (\alpha _{0}+3\alpha _{1})}\,\hat{W}_{i}+%
\frac{1}{2}\,e^{2U}\cos ^{2}\frac{(\alpha _{0}+3\alpha _{1})}{2}\,f_{i}, \\
\partial _{i}\alpha _{1} &=&\frac{1}{2\,\sqrt{3}}\,e^{U}\,\text{Im}\left[
e^{-2i\,\alpha _{1}}\left( 3-2e^{2i\,\alpha _{1}}+e^{i(\,\alpha
_{0}+3\,\alpha _{1})}\right) \hat{\bar{\Pi}}_{i}\right] -  \notag \\
&-&\frac{1}{2}\,e^{U}\,\sin {\ (\alpha _{0}+3\alpha _{1})}\,\hat{W}_{i}-%
\frac{1}{2}\,e^{2U}\cos ^{2}\frac{(\alpha _{0}+3\alpha _{1})}{2}\,f_{i}.
\end{eqnarray}%
As in all other cases, $f_{i}$ is the imaginary part of the quantity whose
real part is $W_{i}$, and it enters both sides of (\ref{almosfal}).

Now, let us try to integrate out electromagnetic potential completely, as
achieved in the composite non-BPS class. By starting from the
Ansatz
\begin{eqnarray}
b^{\beta } &=&\frac{1}{2}\,\text{Re}\left[ \left( 3\,e^{-i\,\alpha
_{0}}-e^{3\,i\,\alpha _{1}}\right) \,v^{\beta }-\sqrt{3}\,e^{-i\,(\alpha
_{0}+2\,\alpha _{1})}\left( 1+e^{i\,(\alpha _{0}+3\,\alpha _{1})}\right)
\,v_{t}^{\beta }\right] e^{U}-  \notag \\
&-&b\,e^{3U}\frac{1}{2}\,\text{Im}\left[ \left( 3\,e^{-i\,\alpha
_{0}}+e^{3\,i\,\alpha _{1}}\right) \,v_{i}^{\beta }-\sqrt{3}\,e^{-i\,(\alpha
_{0}+2\,\alpha _{1})}\left( 1-e^{i\,(\alpha _{0}+3\,\alpha _{1})}\right)
\,v_{t}^{\beta }\right] ,
\end{eqnarray}%
and inserting it into equation (\ref{almosfal}), one achieved the expression
\begin{equation}
\text{Im}\left[ \left( 3\,e^{-i\,\alpha _{0}}+e^{3\,i\,\alpha _{1}}\right) \,Z_{i}-%
\sqrt{3}\,e^{-i\,(\alpha _{0}+2\,\alpha _{1})}\left( 1-e^{i\,(\alpha
_{0}+3\,\alpha _{1})}\right) \,Z_{t\,i}\right] =0.
\end{equation}%
The equations for electromagnetic potential can be solved simultaneously \textit{%
iff} one of these two conditions is fulfilled:
\begin{equation}
\left.
\begin{array}{l}
I:\alpha _{0}+3\,\alpha _{1}=\pi , \\
\\
II:\alpha _{0}+3\,\alpha _{1}=2\arctan (b\,e^{2U}).%
\end{array}%
\right.  \label{alpcond}
\end{equation}%
Correspondingly, two sub-classes $I$ and $II$ of full-fledged almost BPS first order equations exist. 
As we will see in the treatment below,
sub-class $I$ corresponds to BPS-like configurations with a positive
quartic invariant of the sum of all charges pertaining to each center
(namely $I_{4}\left( \sum_{a}Q_{a}\right) >0$ : BPS black hole in the
single-centered limit), whereas sub-class $II$ corresponds to non-BPS
composite-like configurations with a negative quartic invariant of the sum
of all charges pertaining to each center (namely $I_{4}\left(
\sum_{a}Q_{a}\right) <0$ : non-BPS black hole in the single-centered
limit).

The system of flow equations pertaining to sub-class $I$ reads:
\begin{eqnarray}
\partial _{i}U &=&\text{Re}\left[ e^{-i\alpha _{0}}\,Z_{i}\right] e^{U}+\frac{1}{2}%
\,b\,f_{i}\,e^{4U},  \notag \\
\partial _{i}t &=&\frac{\sqrt{3}}{2}\,e^{2i\,\frac{\alpha _{0}-\pi }{3}%
}\left( e^{U}\left( i+be^{2U}\right) f_{i}+2\,i\,\text{Im}\left[ e^{-i\alpha
_{0}}Z_{i}\right] \right) \,e^{U}\,E^{t}+e^{i\,\alpha _{0}}\bar{Z}_{\bar{t}%
\,i}\,e^{U}\,E^{t},  \notag \\
f_{i} &=&\partial _{i}b-2\,e^{-U}\text{Im}\left[ e^{-i\,\alpha _{0}}Z_{i}\right] +%
\sqrt{3}\,b\,e^{U}\,\text{Re}\left[ \sqrt{3}e^{-i\,\alpha _{0}}Z_{i}+i\,e^{-i\,%
\frac{\alpha _{0}-\pi }{3}}Z_{t\,i}\right] ,  \label{falm1}
\end{eqnarray}%
where the phase $\alpha _{0}$ satisfies the condition (already implemented
in the system (\ref{falm1}))
\begin{equation}
\text{Im}\left[ e^{-i\,\alpha _{0}}Z_{i}+i\,\sqrt{3}\,e^{-i\,\frac{\alpha _{0}-\pi
}{3}}Z_{t\,i}\right] =0.  \label{alm1cond}
\end{equation}

{By setting $b=0$ in the system (\ref{falm1}) (and using the condition (\ref%
{alm1cond})), one obtains the BPS first order system (\ref{BPSsyst})
constrained by (\ref{alm1cond}). This leads us to identify this limit of
sub-class }${I}${\ with BPS-like configurations. This conclusion is also
confirmed by the asymptotical analysis, in which the rotation vanishes $%
f_{i}=0$, and $b$ becomes zero (more rapidly than any other functions, due
to nature of rotation); by taking (\ref{alm1cond}) into account, one then
gets a correspondingly constrained BPS system (\ref{BPSsyst}), consistent
with $I_{4}\left( \sum_{a}Q_{a}\right) >0$ in the single-centered
interpretation; this will be considered in one of the examples of the next
section.}

On the other hand, the system of flow equations pertaining to the sub-class $%
II$ reads:
\begin{eqnarray}
\partial _{i}U &=&\frac{1}{2}e^{U}\text{Re}\left[ e^{3i\alpha _{1}}Z_{i}-\sqrt{3}%
e^{i\alpha _{1}}Z_{t\,i}\right] +\frac{1}{2}b\,e^{3U}\text{Im}\left[ e^{3i\alpha
_{1}}Z_{i}-\sqrt{3}e^{i\alpha _{1}}Z_{t\,i}\right] +\frac{1}{2}%
\,b\,f_{i}\,e^{4U},  \notag \\
\partial _{i}t &=&-\frac{1}{2}e^{U}E^{t}\left( \sqrt{3}e^{U-2i\,\alpha
_{1}}\left( i-be^{2U}\right) f_{i}+\sqrt{3}e^{i\,\alpha
_{1}}Z_{i}+e^{-i\,\alpha _{1}}Z_{t\,i}\right) +  \notag \\
&+&\frac{1}{2}e^{U}E^{t}\left( \frac{1}{2}\text{Re}\left[ \sqrt{3}\,e^{3i\alpha
_{1}}Z_{i}-e^{i\alpha _{1}}Z_{t\,i}\right] -\frac{i\,b\,e^{2U-2\,i\,\alpha }%
}{1+b^{2}\,e^{4\,U}}\left( \sqrt{3}\,e^{-3\,i\,\alpha }\bar{Z}%
_{i}-e^{-i\alpha _{1}}\bar{Z}_{\bar{t}\,i}\right) \right) ,  \notag \\
f_{i} &=&\partial _{i}b+\sqrt{3}\,e^{-U}\left( \text{Im}\left[ \sqrt{3}e^{3i\alpha
_{1}}Z_{i}-e^{i\alpha _{1}}Z_{t\,i}\right] -\,b\,e^{2\,U}\text{Re}\left[ \sqrt{3}%
\,e^{3i\alpha _{1}}Z_{i}-e^{i\alpha _{1}}Z_{t\,i}\right] \right) ,
\label{almost-fourth}
\end{eqnarray}%
where the phase $\alpha _{1}$ satisfies the condition (already implemented
in the system (\ref{almost-fourth}))
\begin{eqnarray}
\text{Im}\left[ e^{3i\alpha _{1}}Z_{i}\right]  &=&\frac{\sqrt{3}}{2}\frac{%
b\,e^{2\,U}}{1+b^{2}\,e^{4\,U}}\left( \text{Re}\left[ \sqrt{3}\,e^{3i\alpha
_{1}}Z_{i}-e^{i\alpha _{1}}Z_{t\,i}\right] +b\,e^{2\,U}\text{Im}\left[ \sqrt{3}%
\,e^{3i\alpha _{1}}Z_{i}-e^{i\alpha _{1}}Z_{t\,i}\right] \right) .  \notag
\label{alm2cond} \\
&&  \label{almost-fourth-1}
\end{eqnarray}%
At spatial infinity, {the sub-class }${II}${\ first order system (\ref%
{almost-fourth})-(\ref{almost-fourth-1}) reduces to the non-}${BPS}${\
composite one (\ref{sys-1})-(\ref{sys-2}), when $f_{i}=0$, $b=$constant and
taking (\ref{alm2cond}) into account. For this reason, as mentioned above we
identify this sub-class as a composite non-BPS like one. }

We leave the detailed analysis of the solutions (and related constraints) of
the general almost BPS sub-classes determined above to future
investigation. The next section will be devoted to some illustrative
examples.

\subsection{\label{Ex-2}Some Examples}

\begin{enumerate}
\item As a first example, we consider the single-centered black hole before
integrating the electromagnetic potential out. It should be pointed out that
the almost BPS class admits two possible single-centered limits
(namely, the BPS and non-BPS one), in contrast to the BPS and
composite non-BPS classes. By setting $f_{i}=0$ in (\ref{almost}) and in (%
\ref{almosfal}), and replacing the generalized central charge and its
derivative by the standard ones, one obtains
\begin{equation}  \label{comlimit}
e^{i\,\alpha _{0}}=e^{i\,\alpha _{1}}\,\frac{-\sqrt{3}e^{2i\,\alpha
_{1}}Z+Z_{t}}{-\sqrt{3}\,\bar{Z}+e^{2i\alpha _{1}}\bar{Z}_{\bar{t}}}
\end{equation}%
for a non-BPS single-centered black hole, and
\begin{equation}
e^{i\,\alpha _{1}}\frac{\left( e^{2i\,\alpha _{1}}Z+\sqrt{3}Z_{t}\right) }{%
\bar{Z}+\sqrt{3}e^{2i\,\alpha _{1}}\bar{Z}_{\bar{t}}}=-e^{i\,\alpha _{0}}
\label{cubic}
\end{equation}%
for a BPS single-centered black hole. In particular, (\ref{cubic}) is
cubic in $\alpha _{1}$; its solution is straightforward but cumbersome, and
we will not present it here.

On the other hand, after integrating the electromagnetic fields out, the
single-centered limit yields the following results.{\ }${I)}${\ For case }${I%
}${\ of (\ref{alpcond}), the single-centered non-}${BPS}${\ limit yields a
\textquotedblleft small" black hole. Indeed, from }${I}${\ of (\ref{alpcond}), (\ref{comlimit}%
), (\ref{alm1cond}) and from condition $f_{i}=0$ (where $f_{i}$ is defined
by (\ref{falm1})), the following restrictions on special geometry invariants
\cite{CFMZ1} follow:
\begin{eqnarray}
i_{3} &=&\frac{i}{3\sqrt{3}}\left( Z\bar{Z}_{\bar{t}}^{3}-\bar{Z}%
Z_{t}^{3}\right) =0,  \label{i3=0} \\
i_{2} &=&Z_{t}\bar{Z}_{\bar{t}}=3\,i_{1}=3\,Z\bar{Z},
\end{eqnarray}%
which in turn imply the quartic invariant \cite{CFMZ1}
\begin{equation}
I_{4}=(i_{1}-i_{2})^{2}-4i_{2}^{2}+4i_{4}
\end{equation}%
to vanish. On the other hand, the single-centered }${BPS}${\ limit yields a
\textquotedblleft large" solution; by combining }${I}${\ of (\ref{alpcond}), (\ref{cubic}), (%
\ref{alm1cond}), conditions $f_{i}=0$ and $b=0$, one can indeed check that a
"\textquotedblleft large" BPS black hole solution, constrained by the additional condition (%
\ref{i3=0}) is obtained.}

$II)$ The same procedure applied to {case }${II}${\ of (\ref{alpcond})}
yields no \textquotedblleft large" BPS black holes (all special geometry invariants are
zero), whereas the non-BPS single-centered limit yields the same solutions
as the composite non-BPS class, with no additional constraints.

\item As a second and final example, we consider the well known
almost BPS solution \cite{Goldstein:2008fq}, \cite{Ben}. This is obtained
by constraining the phase $\alpha _{1}$ in the almost BPS
sub-class $II$ (\ref{almost-fourth}) as follows:
\begin{equation}
\alpha _{1}=-\text{arccot}\left[ b\,e^{2U}\right] .  \label{secc}
\end{equation}%
The origin of such a constraint can be better understood by investigating
the differential equations for $\alpha _{0}$ and $\alpha _{1}$, and noticing
that $\alpha _{0}+\alpha _{1}=\pi $ is a particular solution. Combining this
with the second of (\ref{alpcond}), one gets (\ref{secc}). By inserting this
value of $\alpha _{1}$ into (\ref{almost-fourth}) and considering equations
for the combinations $e^{-U}/\sqrt{y}$ and $x+b\,y\,e^{U}$, the following
solutions are obtained:
\begin{eqnarray}
x &=&K_1-\frac{b}{2\,H_{p^{0}}\,Z_{1}},\quad y=\frac{e^{-2U}}{%
2\,H_{p^{0}}\,Z_{1}}, \\
e^{-4\,U} &=&4\,H_{p^{0}}\,Z_{1}^{3}-b^{2},
\end{eqnarray}%
in which the functions $K_1$ (harmonic) and $Z_{1}$ (non-harmonic) respectively
satisfy the equations
\begin{eqnarray}
\partial _{i}K_1 &=&\frac{\partial _{i}H_{p^{0}}\,K_1-\partial _{i}H_{p^{1}}}{%
H_{p^{0}}}, \\
\partial _{i}Z_{1} &=&\frac{1}{3}\partial _{i}H_{q_{1}}+2\partial
_{i}K_1 H_{p^{0}}K_1-\partial _{i}H_{p^{0}}K_1^{2}.
\end{eqnarray}%
As for $f_{i}$, it reads
\begin{equation}
f_{i}=\partial _{i}b-6\,H_{p^{0}}\,Z_{1}\,\partial _{i}K_1,
\end{equation}%
where the harmonic functions are constrained by
\begin{equation}
\partial _{i}H_{q_{0}}=3\,\partial _{i}K_1\,Z_{1}-3\,\partial
_{i}Z_{1}\,K_1+3\,K_1^{2}\,\partial _{i}K_1\,H_{p^{0}}-\partial
_{i}H_{p^{0}}\,K_1^{3},
\end{equation}%
yielded by (\ref{alm2cond}).
\end{enumerate}
We leave the analysis of the physical properties of these solutions, along with an investigation of new classes of solutions, for future work.

\section{\label{Conclusion}Conclusion}

In the present paper, the first order formalism for multi-centered
and/or rotating black holes was constructed and investigated in full
generality for the so-called $T^{3}$ model of $N=2$, $D=4$ ungauged
Maxwell-Einstein supergravity. The exploited procedure sets non-BPS
classes on the same footing as the well known BPS one.

In order to highlight the various steps of our approach to the construction
of the first order flow equations, we first considered the BPS class \cite%
{Denef}.

Similarly to the BPS class, the composite non-BPS class \cite%
{Bossard:2011kz} is endowed with only one auxiliary phase field, which
generally satisfies some algebraic constraint. For a single-centered static
black hole, this phase can be integrated out without putting any restriction
on the charges, and the well known non-BPS fake superpotential for the $%
T^{3}$ model is retrieved. It is worth stressing that a {smooth
single-centered limit is obtained also after integrating electromagnetic
fields out.} For what concerns rotating single-centered solutions, we
recovered a generalization of the Rasheed-Larsen black hole \cite%
{Rasheed:1995zv}, in which all electric and magnetic charges are switched
on. As for multi-centered solutions, we also have retrieved all known
solutions.

Consistent with its nilpotent orbit characterization \cite{Bossard:2011kz},
the almost BPS class exhibits the most involved structure, in
which two phase fields are present. We have shown that two general
sub-classes ($I$ and $II$) exist, depending on two possible constraints
satisfied by the two phases. Also, the almost BPS class
exhibits both a BPS and a non-BPS single-centered (extremal) black hole
limit. In particular, the {BPS single-centered limit yields some
restrictions on the special geometry invariants (explicitly derived in the
discussion of some examples). This is to be contrasted with the BPS and
composite non-BPS classes, which admit only one single-centered
(extremal) black hole limit (BPS and non-BPS, respectively), but without
any restriction on the special geometry invariants.} Furthermore, the
classes $I$ and $II$ of almost BPS flow respectively correspond
to a positive and negative quartic $U$-duality invariant polynomial $I_{4}$,
evaluated on the total charge vector of the system. The well known
almost BPS multi-centered solution was also recovered, and it was shown
that it is nothing but a particular solution of sub-class $II$.

Having derived the most general sets of first order flow equations
(consistent with the flatness of the three dimensional base-space), we leave
the detailed study of their various new classes of solutions, along with
their physical properties, to forthcoming works. The $T^{3}$ model provides
a good framework, due to the presence of only one modulus and the absence of
any flat direction. Furthermore, we plan to extend a{ll
such results to the $STU$ model, which instead exhibits flat directions
\cite{FM-2}.}

\textit{Note added:} the new paper \cite{BosKat} on this subject appeared on arXiv after the present work had been completed.

\section*{Acknowledgements}

{We would like to thank Gianguido Dall'Agata, Sergio Ferrara and Andrey
Shcherbakov for useful discussions. Special thanks to Guillaume Bossard and
to Alessio Marrani for very useful comments
and suggestions. We would like also to thank the organizers of the
Conference \textit{"\textquotedblleft The supersymmetric, the extremal and the ugly -
solutions in string theory"} held in Saclay in November 2011, where
half-baked work was reported, as well as CERN Theory Division, where part of
this work was done. This work is supported by the ERC Advanced Grant no.
226455 SUPERFIELDS.}

\end{document}